\documentclass[twocolumn]{aastex701}
\usepackage{amsmath}

\newcommand\latex{La\TeX}
\begin{document}

\title{Triple Decomposition of Field Gradient Tensors in Space Plasmas}

\author[orcid=0000-0001-7288-1087,sname='Quattrociocchi']{Virgilio Quattrociocchi}
\affiliation{INAF-Istituto di Astrofisica e Planetologia Spaziali, Roma, Italy}
\email[show]{virgilio.quattrociocchi@inaf.it}  

\author[orcid=0000-0002-3403-647X, sname='Consolini']{Giuseppe Consolini} 
\affiliation{INAF-Istituto di Astrofisica e Planetologia Spaziali, Roma, Italy}
\email[]{giuseppe.consolini@inaf.it} 

\author[orcid=0000-0002-5002-6060, sname='Marcucci']{Maria Federica Marcucci} 
\affiliation{INAF-Istituto di Astrofisica e Planetologia Spaziali, Roma, Italy}
\email[]{maria.marcucci@inaf.it} 

\begin{abstract}

Characterizing local topology in magnetohydrodynamic (MHD) turbulence is fundamental to understanding energy dissipation, intermittency, and coherent structures in space plasmas. Classical double decomposition of velocity ($A_{ij}$) and magnetic ($Z_{ij}$) field gradient tensors into symmetric strain and anti-symmetric rotation components conflates rigid-body rotation with planar sliding shear. Here, we extend the real-Schur-based Triple Decomposition framework to MHD flows, rigorously isolating irrotational normal strain ($N, M$), residual pure shear ($H, K$), and true rigid-body rotation ($R, B$). Applying this method to high-resolution NASA-MMS multi-spacecraft observations in the solar wind and Earth's magnetosheath reveals distinct topological signatures. Solar wind dynamics are predominantly strain-dominated due to spherical radial expansion, whereas magnetosheath turbulence is overwhelmingly shear-dominated, driven by boundary interactions. Crucially, cross-term contributions ($2RH$ and $2KB$) are suppressed to $\le 7\%$ across both plasma regimes, contrasting sharply with neutral fluid turbulence ($\sim 13\%$). This suppression is driven by background magnetic-field-induced anisotropy ($k_\perp \gg k_\parallel$), which confines shearing motions perpendicular to the primary rotation axes. By cleanly separating genuine vortex cores from current sheets and shear layers, the Triple Decomposition provides an objective framework for characterizing plasma turbulence and analyzing data from future multi-spacecraft missions.

\end{abstract}

\keywords{\uat{Space plasmas}{1544} ---  \uat{Interplanetary turbulence}{830} --- \uat{Heliosphere}{739} --- \uat{Solar wind}{1534}}


\section{Introduction}
The heliospheric space plasma environment can be seen as a natural laboratory where it is possible to study the principal features of turbulence in magnetized plasma \citep{bruno2016turbulence}.
Indeed, turbulence represents an important characteristic of heliospheric plasmas, manifesting across different regions including the solar wind, the terrestrial central plasma sheet, the Earth's magnetosheath, and the ionosphere. In these contexts, turbulence acts as a primary driver for cross-scale coupling, mediating the transport and exchange of energy between distinct scales down to the kinetic ones where dissipation is expected to occur.

At MHD scales, a nonlinear active turbulent cascade operates in the solar wind. The power-law scaling of fluctuations in the inertial range, spanning periods from hours down to tens of seconds, is approximately of Kolmogorov type, with a spectral index close to $f^{-5/3}$, although the turbulence is also characterized by additional properties such as anisotropy and intermittency. The spectra of magnetic and velocity fluctuations form repeatable, nearly universal observations that extend from the acceleration region near the Sun, as probed by Parker Solar Probe, into the outer heliosphere, as measured by missions such as Ulysses and Voyager. This multi-scale structure of the turbulent cascade has been confirmed to evolve with heliocentric distance: the inertial-range power-law index varies between approximately -1.5 and -1.7 with a weak radial dependence, while the dissipation-range index steepens significantly closer to the Sun, consistent with gyro-resonant damping of fluctuations by thermal protons \citep{lotz2023radial, smith2024unsolved}. Indeed, at higher frequencies, corresponding to sub-ion scales, the power spectrum steepens with a spectral index ranging between -2 and -4 (typically $\sim -2.8$), indicating the onset of kinetic effects where the MHD approximation breaks down and ion kinetic physics becomes relevant. In this range of kinetic dissipation, it is thought that turbulent energy is transferred into a cascade in a nonlinear way. However, the specific dissipation mechanisms and the nature of fluctuations in this regime are still the subject of active investigation. \citep{valentini2010two,howes2014kinetic,cerri2019kinetic, grovselj2019kinetic}.

A distinguishing feature of plasma turbulence from classical neutral-fluid turbulence is its pronounced intermittency \citep{bruno2016turbulence, wan2016intermittency}. The turbulent cascade generates numerous coherent structures—primarily thin current sheets—characterized by intermittent spatial organization, extending from ion to electron scales. These structures are identified as preferential sites of energy dissipation and localized plasma heating \citep{karimabadi2013coherent, wang2021energy}. In the terrestrial magnetosheath, which hosts turbulence amplified from the upstream solar wind and further excited by local plasma instabilities, such coherent structures manifest in a variety of forms and provide an ideal environment to study energy dissipation mechanisms in collisionless plasmas. Multi-spacecraft observations with Cluster have demonstrated that turbulence can spontaneously generate thin proton-scale current sheets in the magnetosheath, which partially account for the intermittency and the spatially inhomogeneous distribution of plasma heating.

Intermittency, which violates the global self-similar and space-filling assumptions of many theoretical turbulence models, is typically quantified through then anomalous scaling of $p-$th order structure functions with respect to the Kolmogorov K41 theory. Magnetic shears associated with coherent structures, such as current sheets, are potential sites where magnetic reconnection can be triggered within turbulent flow, making intermittency a key feature in the context of turbulence-driven reconnection \citep{stawarz2024interplay}.

Thus, the identification and characterization of coherent structures in turbulent flows represent cornerstones of space plasma dynamics research.

The study of some morphological and topological features of turbulent flows dates back to the early works by \citet{vieillefosse1982local} and \citet{cantwell1992exact}, which investigated the features of the velocity gradient tensor. Indeed, for decades, local analysis of fluid motion features has traditionally been dictated by the Cauchy-Stokes theorem, which decomposes the velocity gradient tensor $\nabla \mathbf{v}$, into a symmetric component, the rate-of-strain tensor $\mathbf{S}$, and an antisymmetric component, the vorticity tensor $\mathbf{W}$.
This means that the velocity gradient dynamics and the small scale behavior of turbulence have been studied by decomposing the velocity gradient tensor ($A_{ij} = \partial_j v_i$) into symmetric and anti-symmetric parts:
\begin{equation}
    A_{ij} = S_{ij} + W_{ij}
    \label{Eq1}
\end{equation}
where the symmetric tensor $S_{ij}$ and the skew-symmetric one $W_{ij}$ are defined as follows:
\begin{equation}
    S_{ij} = \frac{1}{2}(A_{ij} + A_{ji)}); \qquad W_{ij}=\frac{1}{2}(A_{ij}-A_{ji})
    \label{Eq2}
\end{equation}
In the same way, we can state that the magnetic field gradient tensor $Z_{ij}=\partial_j b_i$ can be decomposed into its symmetric and skew-symmetric contributions:
\begin{equation}
    Z_{ij} = O_{ij} + U_{ij}
    \label{Eq3}
\end{equation}
with
\begin{equation}
    O_{ij} = \frac{1}{2}(Z_{ij}+Z_{ji}); \qquad U_{ij} = \frac{1}{2}(Z_{ij}-Z_{ji})
    \label{Eq4}
\end{equation}

This decomposition is very useful to obtain fundamental insight on small-scale intermittency, intense structures in turbulence, and about the local geometry/topology of stream and magnetic lines. The symmetric contribution of the original tensors ($S_{ij}$ and $O_{ij}$) is related to the strain rate tensor, while the skew-symmetric contribution ($W_{ij}$ and $U_{ij}$) to the rotation rate tensor. Although this framework is mathematically elegant, its physical interpretation in complex turbulent environments is often problematic. Specifically, the skew-symmetric tensors, $\mathbf{W}$ and $\mathbf{U}$ for velocity and magnetic field, respectively, mix up the effects of rigid-body rotation with those of local shearing motion. This kind of contamination makes it difficult to distinguish between the actual cores of vortices and regions of intense shear, leading to potential inaccuracies in defining the topology and the evolution of turbulent structures. Hydrodynamic studies have evidenced that the strain-rate and vorticity do not identify the presence of normal-straining and rigid body rotation of the fluid. To address these conceptual ambiguities, \citet{das2020revisiting} proposed a more refined approach known as the Triple Decomposition of the velocity gradient tensor based on the Kolar's description \citep{kolar2009compressibility}, of the relative motions in which it is possible to find an explanation of how shear can affect both strain and rotation in the double decomposition. By introducing a rigorous mathematical procedure to isolate the contribution of \textit{pure shear} from the classical strain and rotation terms, the triple decomposition of the gradient tensor allows for a partition of the gradient tensor into three kinematically distinct parts: a purely irrotational stretching/compression term, a rigid-body rotation term, and a residual shear term. In a fluid context, the triple decomposition was introduced by extracting the pure-shear contribution from the vorticity, and this technique has been used to identify vortex-structures and to investigate internal shear layers in well-bounded flows.

In the framework of magnetohydrodynamics (MHD), the necessity for such triple decomposition becomes even more important. The dynamics of magnetized fluids are governed by the non-linear interplay between the velocity field $\vec{\mathbf{v}}$ and the magnetic field $\vec{\mathbf{B}}$, where the induction equation couples the motion of the plasma to the evolution of the magnetic topology. Similarly, as the velocity gradient dictates the flow topology, the magnetic field gradient tensor $\nabla \mathbf{B}$, is fundamental in determining the morphology of the magnetic configuration. Indeed, the magnetic field gradient tensor governs critical phenomena such as magnetic reconnection, the formation of current sheets, and the amplification of fields through dynamo processes.

This study investigates the potential of Triple Decomposition for characterizing the morphological and topological features of streamlines and magnetic field lines in turbulent space plasmas. To achieve this, we apply the methodology to velocity and magnetic gradient tensors derived from four solar wind and magnetosheath data intervals observed by NASA's MMS constellation. By analyzing the three components, rotation, stretching, and shear for both fields, our aim is to discover how these fundamental modes of deformation correlate across different scales and regimes. Our goal is to combine the Schur decomposition of gradient tensors in normal and non-normal contributions and a decomposition based on the `Rortex' concept \citep{liu2018rortex}. The last decomposition can separate the rigid body rotation from shear in vorticity. In other words, we can say that we want to examine gradient tensors statistics in MHD turbulence using the decomposition of the tensors into normal-strain rate, rigid body rotation, and pure-shear tensors.\\
\newline
The introduced triple decomposition can improve our understanding of many aspects of gradient dynamics in turbulent magnetized flows:
\begin{itemize}
    \item local structures of flow fields (velocity and magnetic field) in regions characterized by intermittency;
    \item clear separation of solid-body rotation and vorticity;
    \item relation between pressure and gradient tensor fluctuations
\end{itemize}

This work aims to present a novel method for investigating MHD turbulence across various regions of circumterrestrial space by exploiting data collected from multi-spacecraft missions, as the past ESA-CLUSTER and NASA-MMS, or future ones as ESA-Plasma Observatory and NASA-HelioSwarm. To show the potentiality of the present decomposition we apply it to a set of 4 solar wind and Earth's magnetosheath crossing of the NASA-MMS mission \citep{quattrociocchi2026schur}.

\section{On the Triple Decomposition: A brief Resume'}

The core of our diagnostic approach lies in the extension of the Triple Decomposition of a gradient tensor, originally formulated for the velocity gradient tensor, to the magnetic field environment. Following the methodology proposed by \citet{das2020revisiting}, we decompose the velocity gradient tensor $\nabla \mathbf{v}$ and the magnetic field gradient tensor $\nabla \mathbf{B}$ into three physically distinct components. For each field, the gradient tensor is decomposed into a purely irrotational stretching/compression part, a rigid-body rotation part, and a residual pure shear part. This decomposition allows us to move beyond the classical symmetric-antisymmetric partition by isolating the shearing motion, which otherwise contaminates the vorticity and magnetic rotation terms.\\
\newline
The additive decomposition of the velocity field gradient tensor ($\mathbf{A}$) is into a normal strain rate tensor ($\mathbf{N}$), a pure shear tensor ($\mathbf{H}$) and a rigid body rotation rate tensor ($\mathbf{R}$) and is given by
\begin{equation}
    A_{ij} = N_{ij} + H_{ij} + R_{ij}
    \label{Eq5}
\end{equation}
For the magnetic field gradient tensor we can write a similar expression:
\begin{equation}
    Z_{ij} = M_{ij} + K_{ij} + B_{ij}
    \label{Eq6}
\end{equation}
In the last equation, the vectorial meaning of the involved terms is the same as in the velocity case, but the physical interpretation of them is obviously different.\\
\newline
The triple decomposition of velocity and magnetic field gradient tensors can be seen as a natural evolution of the Schur decomposition, as \cite{keylock2018schur} in the hydrodynamic case and \cite{quattrociocchi2026schur} for the magnetohydrodynamic one, of the same matrices. From a practical point of view, let $\mathbf{M}$ be a $3\times3$ matrix representing a generic gradient tensor, then we can perform the triple decomposition of \textbf{M} in the following way:
\begin{itemize}
    \item Perform the real Schur decomposition \citep{schur1909charakteristischen, keylock2018schur} of $\mathbf{M}$ such that $\mathbf{M} = \mathbf{Q}\mathbf{T}\mathbf{Q}^T$. Let $\mathbf{T} = \mathbf{N'} + \mathbf{\Lambda'}$, where $\mathbf{N'}$ is a diagonal matrix containing the real parts of the eigenvalues of $\mathbf{M}$, and $\mathbf{\Lambda'}$ is a strictly upper triangular matrix.
    \item $\mathbf{N'}$ represents the first term of the triple decomposition in the Schur reference frame (normal strain).
    \item Identify whether the matrix $\mathbf{T}$ contains $2\times2$ blocks on the diagonal and, if so, calculate their antisymmetric part $\mathbf{\Omega}$.
    \item $\mathbf{\Omega}$ is the second term of the triple decomposition, representing the solid-body rotation in the Schur reference frame.
    \item Calculate the pure shear term in the Schur frame as the remainder: $\mathbf{H'} = \mathbf{T} - \mathbf{N'} - \mathbf{\Omega}$.
    \item Transform the terms back to the original reference frame: $\mathbf{N}=\mathbf{Q}\mathbf{N'}\mathbf{Q}^T$, $\mathbf{R}=\mathbf{Q}\mathbf{\Omega}\mathbf{Q}^T$, and $\mathbf{H}=\mathbf{Q}\mathbf{H'}\mathbf{Q}^T$.
    \item Verify that the triple decomposition $\mathbf{M} = \mathbf{N} + \mathbf{R} + \mathbf{H}$ holds.
\end{itemize}

\begin{itemize}
    \item perform the real Schur decomposition \cite{schur1909charakteristischen} and \cite{keylock2018schur}, of $\mathbf{M}$, call it $\mathbf{T} = \mathbf{N'} + \mathbf{\Lambda'}$ where $\mathbf{N'}$ is a diagonal matrix containing the eigenvalues of \textbf{M} in an arbitrary order and $\mathbf{\Lambda'}$ is a strictly upper/lower triangular matrix.
    \item $\mathbf{N'}$ is the first term of the triple decomposition in the new reference frame, the normal strain in the new reference frame
    \item identify whether the matrix $\mathbf{T}$ contains $2\times2$ rotation blocks and, if so, calculate their antisymmetric part $\mathbf{\Omega}$
    \item $\mathbf{\Omega}$ is the second term of the triple decomposition in the new reference frame, the solid-body rotation in the Schur-reference frame
    \item calculate $\mathbf{H'} = \mathbf{T} - \mathbf{N'} - \mathbf{\Omega}$
    \item $\mathbf{H'}$ is the last term in the triple decomposition in the new reference frame, the pure shear in the Schur-reference frame
    \item come back to the original reference frame with an anti-Schur transformation: $\mathbf{N}=\mathbf{Q}\mathbf{N'}\mathbf{Q}^T$, $\mathbf{H}=\mathbf{Q}\mathbf{H'}\mathbf{Q}^T$ and $\mathbf{R}=\mathbf{Q}\mathbf{\Omega}\mathbf{Q}^T$
    \item check that the expression $\mathbf{M} = \mathbf{N} + \mathbf{H} + \mathbf{R}$ is true
\end{itemize}

For the purpose of this work, the last two steps are not strictly necessary as we only calculate the Frobenius norm of each tensor, and this quantity is invariant under orthogonal rotations ($\|\mathbf{A}\|_F = \|\mathbf{Q}\mathbf{A}\mathbf{Q}^T\|_F$). For a generic square matrix, its Schur form can be computed as follows:
\begin{equation}
    \mathbf{T} = \mathbf{Q}^T\mathbf{M}\mathbf{Q}
\end{equation}
where $\mathbf{Q}$ is an orthogonal rotational matrix such as $\mathbf{Q}\mathbf{Q}^T = \mathbf{I}$, where $\mathbf{I}$ is the identity matrix. In the most general case, we expect the matrix $\mathbf{T}$ to take the following form:

\begin{equation}
\mathbf{T} =
\begin{bmatrix}
\lambda_{cr} & -\phi & 0 \\
\phi +s_3 & \lambda_{cr} & 0 \\
s_2 & s_1 & \lambda_r
\end{bmatrix}
\label{Eq8}
\end{equation}

The term $-\phi$, the only one not in the triangular matrix, is present if and only if the original matrix has complex eigenvalues, so we can say that this element will be present only if the original vector field shows a rotational contribution and this is a peculiarity of the real Schur transformation. In Eq.(\ref{Eq8}), $\lambda_{cr}$ is the real part of the complex conjugate eigenvalues and $\lambda_r$ is the only real eigenvalue of the starting matrix $\mathbf{M}$. Performing the triple decomposition:
\begin{equation}
    \mathbf{T} = \mathbf{N'} + \mathbf{H'} + \mathbf{R'}
\end{equation}
the three new matrices take the following form:
\begin{eqnarray}
\nonumber \mathbf{N'} &=& \begin{bmatrix}
\lambda_{cr} & 0 & 0 \\
0 & \lambda_{cr} & 0 \\
0 & 0 & \lambda_r
\end{bmatrix} 
\\
\mathbf{H'} &=& \begin{bmatrix}
0 &\  \ 0 &\  \ 0 \\
s_3 &\  \ 0 &\  \ 0 \\
s_2 &\  \ s_1 &\  \ 0
\end{bmatrix}
\\
\nonumber \mathbf{R'} &=& \begin{bmatrix}
0 &\ \ -\phi &\ \ 0 \\
\phi &\ \ 0 &\ \ 0 \\
0 &\ \ 0 &\ \ 0
\end{bmatrix}
\label{Eq10}
\end{eqnarray}
The tensor $\mathbf{H'}$ represents the pure shearing deformation of the fluid/magnetic element, it is a lower/upper triangular tensor with three elements ($s_1$, $s_2$ and $s_3$) that constitute the transverse gradients of velocity or magnetic field components. On the other hand, the tensor $\mathbf{R'}$ is an anti-symmetric tensor that represents pure rotation of the fluid/magnetic element with an angular velocity/current density given by $\phi$. All primed quantities are calculated in the new (Schur) reference frame, to obtain the expressions in the original reference frame we have to come back using the anti-Schur transformation: $\mathbf{W} = \mathbf{Q}\mathbf{T}\mathbf{Q}^T$.

Summarizing the triple decomposition allows to decompose the velocity gradient tensor and the magnetic gradient tensor into three main contribution which are associated to different features of the field lines. In detail, for the velocity field we have:
\begin{itemize}
    \item[i)] $\mathbf N$: the normal-strain-rate tensor that represents pure stretching/compression of a fluid element along the local principal axes. It is the \textit{irrotational straining} term. This tensor term is non-normal, being the off-diagonal part of the local decomposition;
    \item[ii)] $\mathbf H$: the pure-shear tensor that is associated with the motion where fluid layers slide past one another;
    \item[iii)] $\mathbf R$: the rigid-body-rotation-rate tensor which represents the genuine solid-body-like rotation of the fluid element. This term is antisymmetric and isolates only the part of local spinning that is truly coherent, i.e., the rigid rotation.
\end{itemize}

Analogously, for the magnetic field tensor we can assume that:
\begin{itemize}
    \item[i)] $\mathbf M$: the normal-strain-rate analog. This is a measure of the magnetic field-line stretching and compression in the local principal directions, i.e., the rate at which flux tubes are being elongated or squeezed. This term is related to magnetic energy conversion. Indeed, field-line stretching against magnetic tension is exactly the mechanism behind the stretch-twist-fold picture of dynamo action and local magnetic energy amplification;
    \item[ii)] $\mathbf K$: the shear analog, which measures the magnetic shear. This is the tensor quantity most directly related to current sheets. It is important to underline that the squared-Fr\"obenius norm $||\mathbf K||_F^2$ is a natural candidate for a {\it local, frame-consistent current-sheet intensity metric};
    \item[iii)] $\mathbf B$: the rigid-rotation analog. It provides a measure of local field-line rotation without deformation. Rather than shear-driven rotation from a current sheet, this would physically correlate to something like a {\it local flux-rope or magnetic-vortex-like twist} in the field-line geometry.
\end{itemize}
It is important to emphasize that the analogy between the decomposition terms of the velocity gradient tensor and magnetic field one is just an our extrapolation based on the general vector field gradient tensor decomposition and considering the physical meaning associated with the magnetic field line deformation in MHD.

\section{Data Description and Methodology}

Data used for this study are from two different plasma environments: the solar wind and Earth's magnetosheath, and they are measured by the NASA-MMS space mission. This mission consists of four satellites flying in a tetrahedral (tetrad) configuration, which makes it possible to compute spatial gradients directly from simultaneous multi-point measurements. The studied regions provide a natural laboratory for studying magnetohydrodynamic turbulence across different regimes of plasma beta and compressibility. The solar wind data represent a regime of developed, nearly incompressible turbulence, while the magnetosheath data offer insights into a more compressed and turbulent environment downstream of the bow shock. The principal properties of the investigated periods are summarized in Table \ref{tab1} (for a more detailed description of the studied periods, refer to \cite{chhiber2018higher}, \cite{bandyopadhyay2020situ}, and \cite{bandyopadhyay2020statistics}).

\begin{table}[]
    \centering
    \begin{tabular}{c|c|c|c|c|c}
         \# & Data & Time & $\langle\mathbf B\rangle$& $d_i$ & $v$\\
        & (dd/mo/yr) & (UT) & (nT) & (km) & ($\frac{km}{s}$) \\
        \\
        \hline
        SW1 & 24/11/17 & 01:10-02:10 & 6.6 & 76 & 377 \\
        SW2 & 26/11/17 & 21:09-22:09 & 7.7 & 75 & 330 \\
        MSH1 & 26/12/17 & 06:12-06:52 & 17.9 & 48 & 238 \\
        MSH2 & 27/10/18 & 09:13-09:58 & 8.6 & 56 & 278
    \end{tabular}
    \caption{General features of the selected solar wind and magnetosheath time intervals of MMS measurements: $\langle\mathbf B\rangle$ = mean magnetic field; $d_i$ = ion inertial length; $v$ = mean velocity. MSH and SW stand for magnetosheath and solar wind, respectively.}
    \label{tab1}
\end{table}

We consider high-resolution magnetic field and velocity field measurements in the GSE reference system for all four MMS satellites, together with the corresponding orbital parameters, i.e. the satellite positions in the GSE reference system. The original data time resolution is 128 samples/s for the magnetic field and 0.15 s ($\sim$6.7 samples/s) for the velocity (ion) field. Velocity field measurements were corrected for spectral spikes using the Hampel filtering method described in \cite{bandyopadhyay2018solar}, which identifies and replaces outliers falling outside a robust, median-based threshold within a moving window, thus preserving genuine turbulent fluctuations while removing instrumental artifacts.

For each of the selected intervals, all quantities were linearly interpolated onto a common time grid at the cadence of the coarser-resolution dataset (the velocity field), to ensure that the magnetic field, velocity field, and position data are synchronized across the four spacecraft at every time step used in the gradient computation.

\subsection{Methods: The Tetrad Approach}

\begin{figure}
    \centering
    \includegraphics[width=0.5\linewidth]{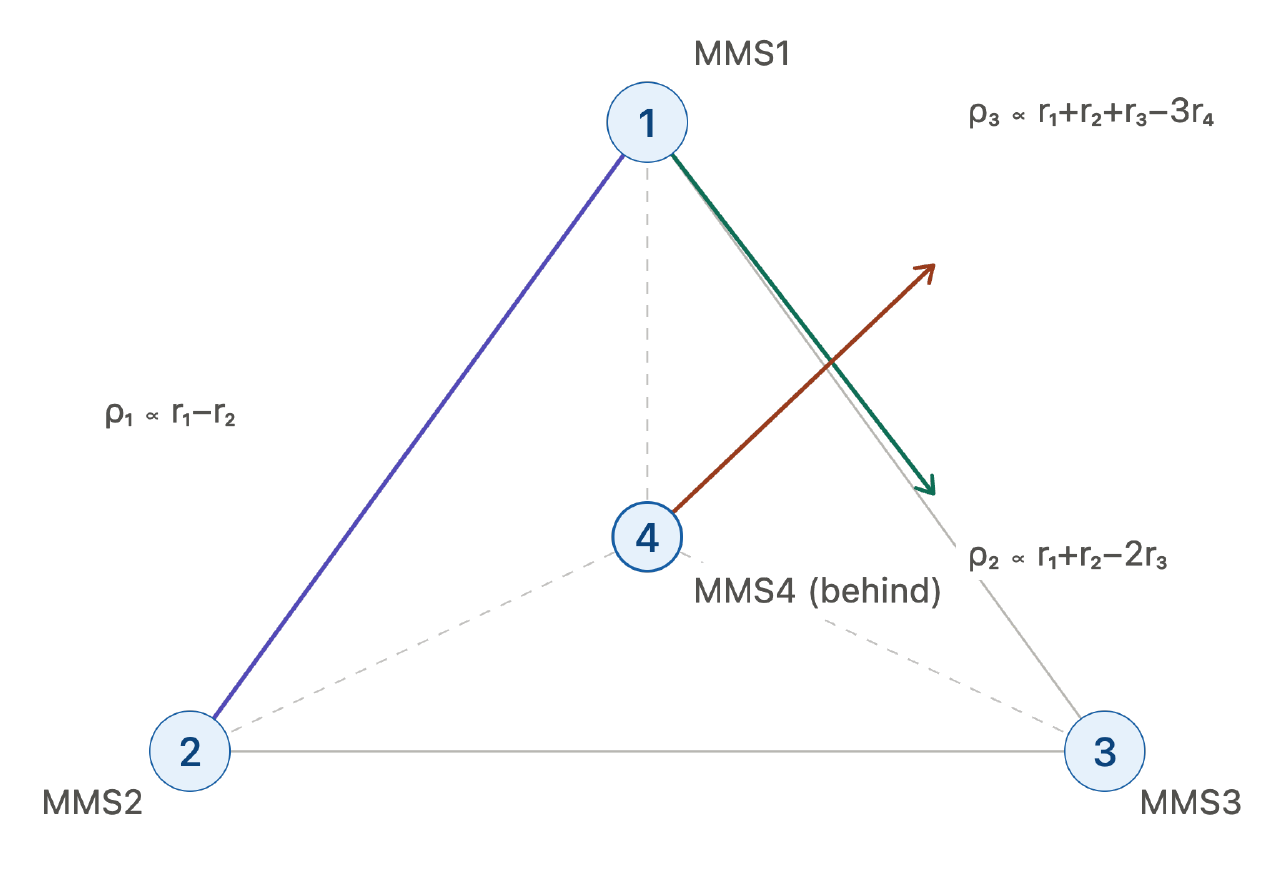}
    \caption{A schematic picture for the construction of the distance matrix $\hat\rho$ using the Tetrad method. Credit: Generated with the assistance of Claude Sonnet 5 (Anthropic, July 2026).}
    \label{fig:01}
\end{figure}

We compute the spatial gradient of the magnetic field and of the velocity field at each time step using the four-point technique developed for tetrahedral configurations \citep{chanteur1998spatial, chertkov1999lagrangian, consolini2015statistics, quattrociocchi2019geometrical}. This method, named \textit{Tetrad method}, is based on the Chanteur-type formalism (orthogonal "natural coordinate" combinations of the four spacecraft positions), which is more specific than the generic reciprocal-vector description in the case of tetrahedral configurations \citep{chanteur1998spatial}. For each time step, we first construct, from the position vectors $\mathbf r_1, \mathbf r_2, \mathbf r_3, \mathbf r_4$ of the four MMS spacecraft (in the barycentric frame), three mutually orthogonal relative-position vectors,
\begin{align}
    \boldsymbol\rho_1 &= \frac{(\mathbf r_1 - \mathbf r_2)}{\sqrt{2}}, \\
    \boldsymbol\rho_2 &= \frac{(\mathbf r_1 + \mathbf r_2 - 2\mathbf r_3)}{\sqrt{6}}, \\
    \boldsymbol\rho_3 &= \frac{(\mathbf r_1 + \mathbf r_2 + \mathbf r_3 - 3\mathbf r_4)}{\sqrt{12}} \, ,
\end{align}
which encode the relative geometry of the tetrahedron independently of its absolute position. Figure \ref{fig:01} shows the construction of the distance matrix $\hat \rho$ using the Tetrad method. The identical linear combination is applied to the field measured at each spacecraft, $\mathbf f_1,\dots,\mathbf f_4$ (i.e. $\mathbf B_\alpha$ or the plasma velocity, depending on the field under analysis), to obtain the corresponding field-difference vectors,
\begin{align}
    \mathbf u_1 &= \frac{(\mathbf f_1 - \mathbf f_2)}{\sqrt{2}}, \\
    \mathbf u_2 &= \frac{(\mathbf f_1 + \mathbf f_2 - 2\mathbf f_3)}{\sqrt{6}}, \\
    \mathbf u_3 &= \frac{(\mathbf f_1 + \mathbf f_2 + \mathbf f_3 - 3\mathbf f_4)}{\sqrt{12}} \, .
\end{align}

Denoting by $\hat\rho$ the $3\times3$ matrix whose rows are $\boldsymbol\rho_1,\boldsymbol\rho_2,\boldsymbol\rho_3$, and by $\hat u$ the analogous matrix built from $\mathbf u_1,\mathbf u_2,\mathbf u_3$, the gradient tensor $M_{ij} \equiv \partial_i F_j$ of the field $\mathbf F$ is obtained as
\begin{equation}
    M_{ij} = \left(\hat\rho^{-1}\right)_{ik} u_{kj} - \frac{\delta_{ij}}{3}\,\mathrm{Tr}\!\left(\hat\rho^{-1}\hat u\right) \, .
\end{equation}

The second term on the right-hand side explicitly removes the trace of the tensor by construction, so the resulting $M_{ij}$ is traceless independently of measurement noise or of small deviations from an ideal tetrahedron geometry. Applying this expression to the magnetic field yields the tensor $\mathbf Z \equiv \nabla\mathbf B$, whose tracelessness enforces $\nabla\cdot\mathbf B = 0$ exactly, consistently with the physical divergence-free constraint. This procedure removes the spurious divergence while preserving the symmetric and antisymmetric structure of the tensor, from which the rate-of-strain and current density (via $\mathbf J = \nabla\times\mathbf B/\mu_0$) can subsequently be derived.

The same traceless projection was applied to the velocity gradient tensor $\mathbf A$, under the working assumption of local incompressibility ($\nabla\cdot\mathbf v \simeq 0$). This assumption is supported by the fact that the trace of the originally computed $\mathbf A$ was found to be small compared to the other invariants of the tensor (e.g. relative to its Frobenius norm) across the analyzed intervals, indicating that compressive effects are subdominant with respect to the incompressible, solenoidal component of the flow on the scales probed by the tetrad. We note that this treatment is best justified for the solar wind intervals, where compressibility is intrinsically weak, whereas for the magnetosheath intervals it should be regarded as an approximation, given the higher plasma beta and the larger compressive fluctuations typically observed downstream of the bow shock.

As for the magnetic field case, the reliability of this technique depends on the quality of the instantaneous tetrahedral configuration, commonly quantified through its planarity and elongation \citep{consolini2015statistics}. All intervals selected for this study (Table \ref{tab1}) were chosen because the MMS tetrad maintained a nearly regular geometry throughout, minimizing the geometric errors intrinsic to this gradient estimation technique.

\subsection{Methods: Fr\"obenius Norm}

To establish the relevance of the three terms of the triple decomposition of the gradient tensors, we use of the Fr\"obenius norm of a matrix.

The Fr\"obenius norm (also named the Schur norm) of a $m\times n$ matrix $\mathbf A$ 
is a natural way to measure the ``size" or ``magnitude" of a matrix, and it
is defined as 
\begin{equation}
    ||\mathbf A||_F = \sqrt{\sum_{i=1}^m\sum_{j=1}^n a_{i,j}^2}=\sqrt{\mathrm{Tr}(\mathbf A^*\mathbf A)}
    \label{Frobenius}
\end{equation}
where $\mathbf A^*$ is the conjugate transposed matrix. 

 If $\mathbf Q$ is an orthogonal matrix, i.e., $\mathbf Q^T\mathbf Q = \mathbf Q\mathbf Q^T=\mathbf I$ where $\mathbf I$ is the identity matrix, the Fr\"obenius norm of $\mathbf N =\mathbf Q^T\mathbf N'\mathbf Q$ is conserved, i.e.,
\begin{equation}
    ||\mathbf N||_F = || \mathbf Q^T\mathbf N'\mathbf Q||_F = ||\mathbf N'||,
    \label{normconservation}
\end{equation}
which is the case of the unitary Schur transformation. Thus, we will see that we can directly compute the magnitude of the three terms of the triple decomposition directly on the matrix $\mathbf N'$, $\mathbf H'$ and $\mathbf R'$.

An important issue to keep in mind is that because the triple decomposition is not Pythagorean, i.e., the three matrix are not orthogonal, the sum of the Fr\"obenius norms of the three matrix term is not equal to the Fr\"obenius norm of the gradient tensor. For example, in the case of the velocity gradient tensor $\mathbf A = \mathbf N + \mathbf H+\mathbf R$, being $\mathbf H$ and $\mathbf R$ not orthogonal, we have

\begin{equation}
    ||\mathbf A||_F^2 \neq ||\mathbf N||_F^2+||\mathbf H||_F^2+||\mathbf R||_F^2.
    \label{normdifference}
\end{equation}
Indeed, $A_{ij}A_{ij}= N_{ij}N_{ij}+H_{ij}H_{ij}+R_{ij}{ij}+2R_{ij}H_{ij}$, so that there is an extra term $2\mathbf R\mathbf H = 2\phi s_3$ to be considered.
This term quantifies tells how much of a vortex's low-pressure core is due to true rigid rotation versus shear-augmented rotation.

Using the Fr\"obenius norm, we can introduce a measure of the relative contribution of each term of the decomposition by the following expression
\begin{equation}
    w^{...}_F = \frac{||...||_F^2}{||\mathbf N||_F^2+||\mathbf H||_F^2+||\mathbf R||_F^2+2||\mathbf R\mathbf H||_F}
    \label{normweight}
\end{equation}
where $||...||_F^2$ is the Frobenius-norm-squared.

\section{Results}

We begin our study of the triple decomposition by first comparing it with the traditional decomposition based on the symmetric and antisymmetric parts. This comparison is carried out by computing the Fr\"obenius norm of the symmetric and antisymmetric parts of the gradient tensor, as well as of the three contributions obtained from the triple decomposition, and by plotting these values on the $(R,Q)$-plane.  

\begin{figure}
    \centering
    \includegraphics[width=0.75\linewidth]{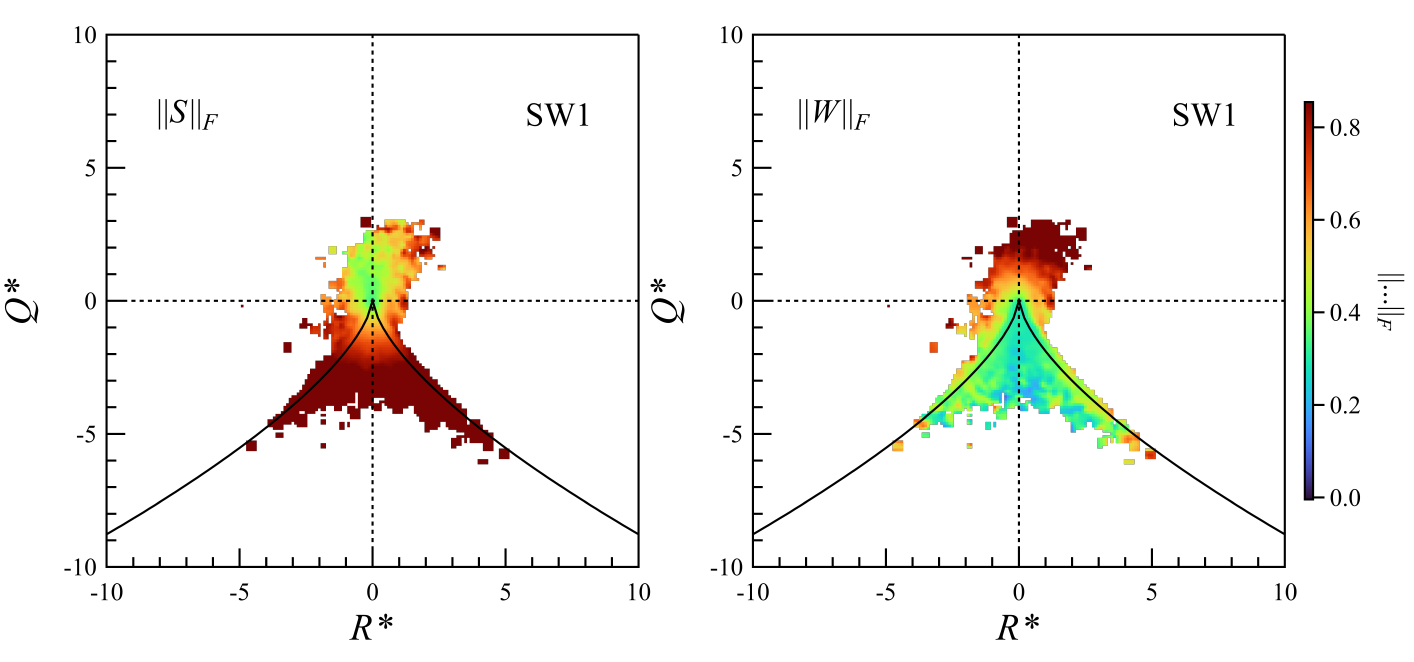}\\
    \includegraphics[width=0.75\linewidth]{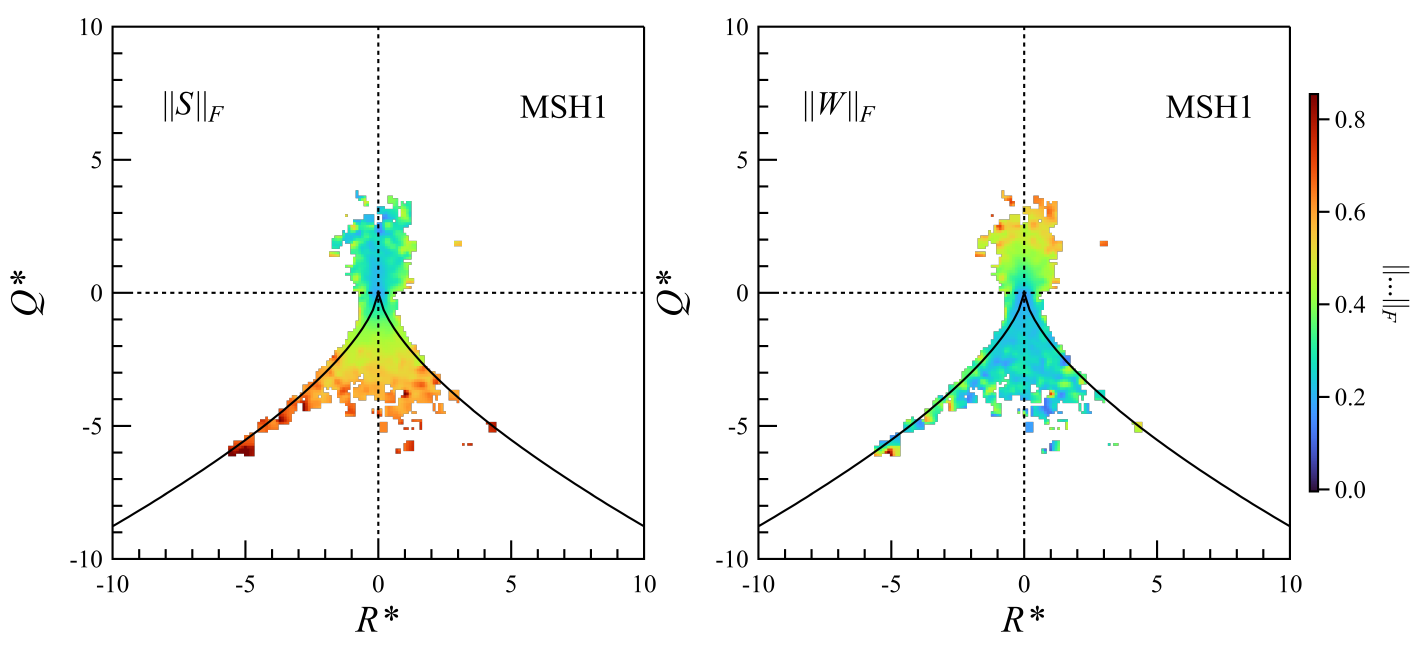}
    \caption{Distributions of the values of the Fr\"obenius norm of the symmetric $\mathbf S$ and antisymmetric $\mathbf W$ parts of the velocity field gradient tensor for the first solar wind period (SW1) and the first magnetosheath one (MSH1). Here, $^*$ indicates standard deviation scaled invariants: $Q^* = Q/\sigma_Q$ and $R^* = R/\sigma_Q^\frac{3}{2}$}
    \label{fig:02}
\end{figure}

Figure \ref{fig:02} shows the distributions of the values of the Fr\"obenius norm of the symmetric and antisymmetric parts of the velocity field gradient tensor on the $(R,Q)$-plane for SW1 and MSH1 intervals. As expected, while the antisymmetric part, which is associated with the vorticity tends to localize the higher values of the norm in the region above the discriminant line $\Delta$, defined as follows:
\begin{equation}
    \Delta = \frac{Q^3}{27} + \frac{R^2}{4},
    \label{discrimin}
\end{equation}
the symmetric part is higher below the discriminant line.  

\begin{figure}
    \centering
    \includegraphics[width=0.99\linewidth]{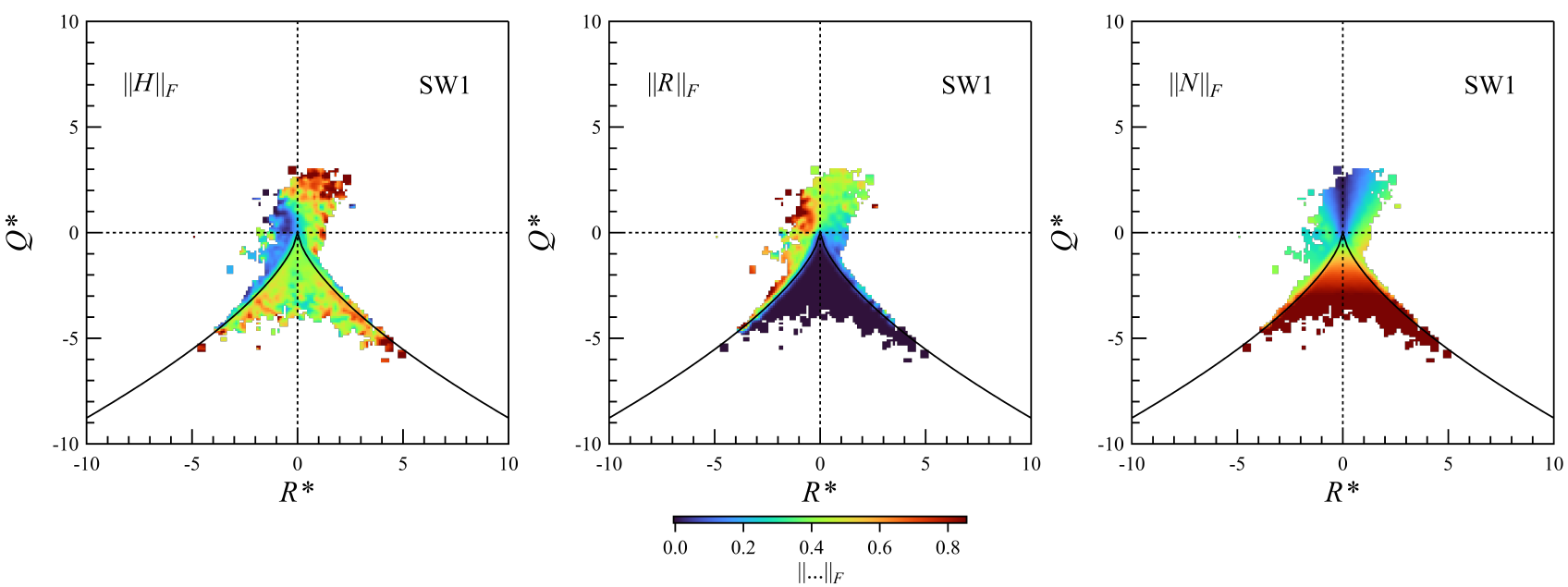}\\
    \includegraphics[width=0.99\linewidth]{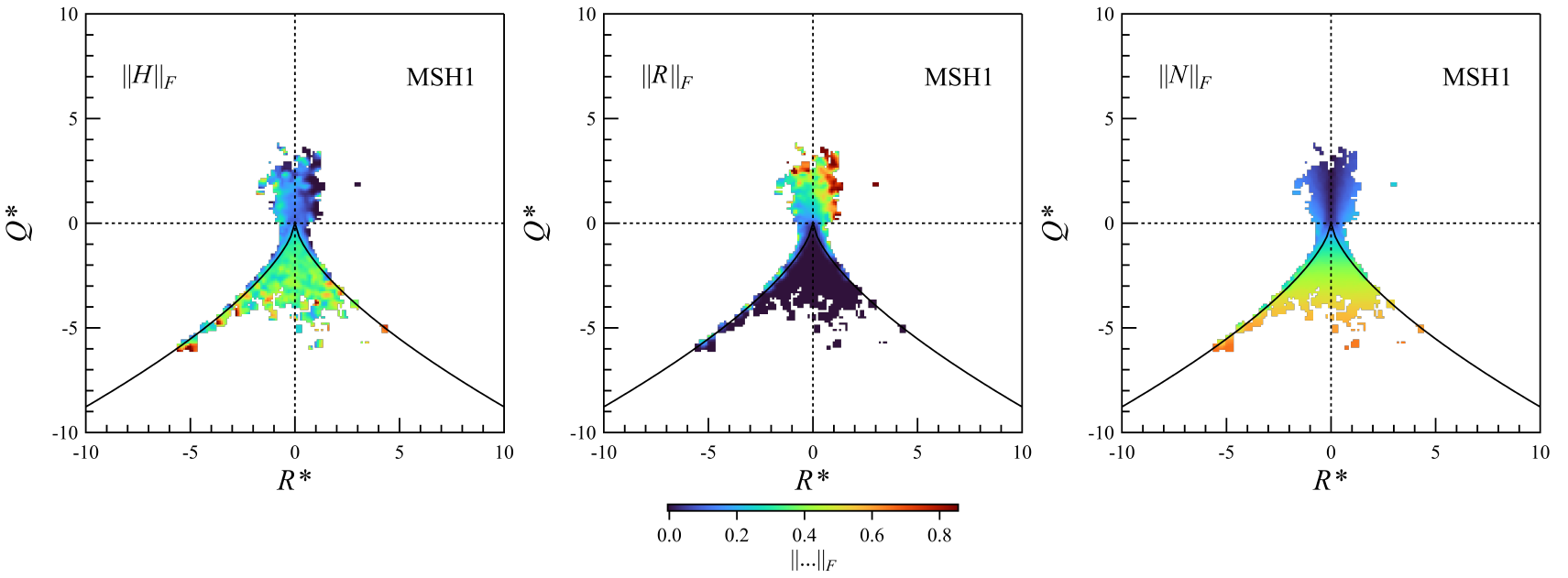}
    \caption{Distributions of the values of the Fr\"obenius norm of the three components of the triple decomposition of the velocity field gradient tensor for the first solar wind period (SW1) and the first magnetosheath interval (MSH1).}
    \label{fig:03}
\end{figure}

Differently, the triple decomposition (see Figure \ref{fig:03}) seems to be  capable to better characterize the different structure. Indeed, the distribution of the Fr\"obenius norms for the rigid body rotation term $\mathbf R$ and the elongation/stretching one $\mathbf N$ are characterized by values significantly different from zero above and below the discriminant line, respectively. This agrees with the rotational character of the flow (mainly incoming and outcoming vortices) above the discriminant line and with the tube-like/sheet-like features of streamlines below $\Delta =0$. For what regards the Fr\"obenius norm of the $\mathbf H$ term, which is mainly associated with the shear character of the streamline, its distributions does not show a well defined location on the $(R,Q)$ plane. The only notable feature seems to be its increase moving away from the center of the plane.

The next step of our analysis lies in investigating the relative contribution of shear, elongation/stretching, and solid-body rotation in each of the selected intervals. In this way, we are able to describe three different pure regimes in the motion and are able to describe the fluid motion by them, the standard decomposition allows us to investigate the vector field structures from a topological point of view \cite{chong1990general}, and using the triple decomposition it is possible to talk about how important a specific contribution to the magnetic fluid motion is. These quantities are evaluated by computing the corresponding Frobenius' norm for each term and successively evaluating the quantity $w_F^{...}$ defined in Eq. \ref{normweight}.  

\begin{figure}
    \centering
    \includegraphics[width=0.9\linewidth]{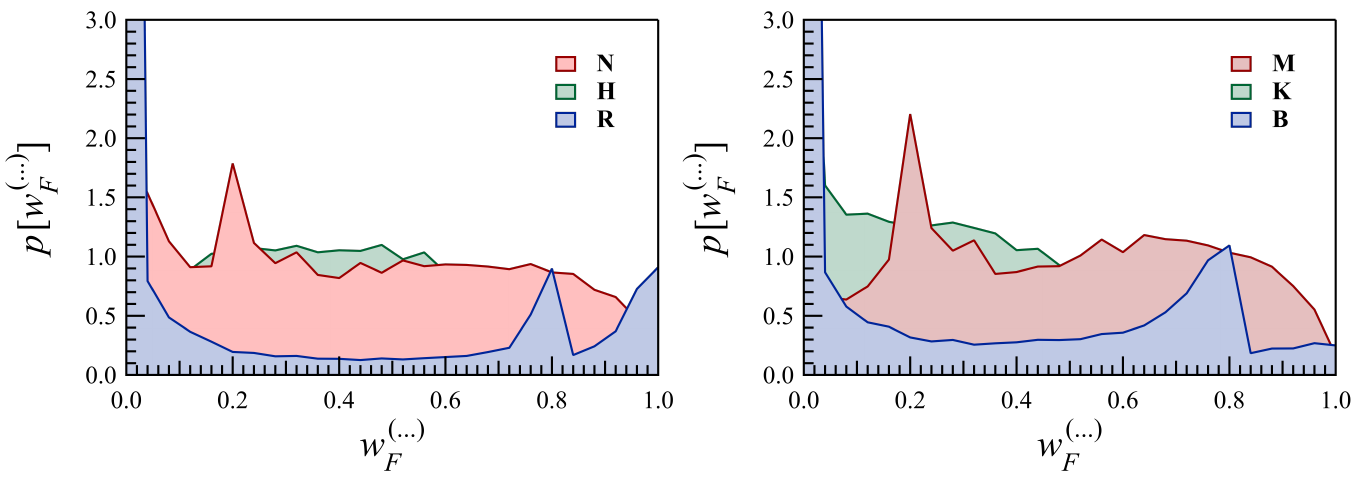}
    \caption{Probability density function (PDFs) of the relative weights $w_F^{(...)}$ for the three components ($\mathbf N$ and $\mathbf M$ elongation/stretching; $\mathbf H$ and $\mathbf K$ shear; $\mathbf R$ and $\mathbf B$ rigid body rotation)  of the triple decomposition of the velocity field (left panel) and magnetic field (right panel) gradient tensors for the first solar wind period (SW1).}
    \label{fig:04}
\end{figure}

\begin{figure}
    \centering
    \includegraphics[width=0.9\linewidth]{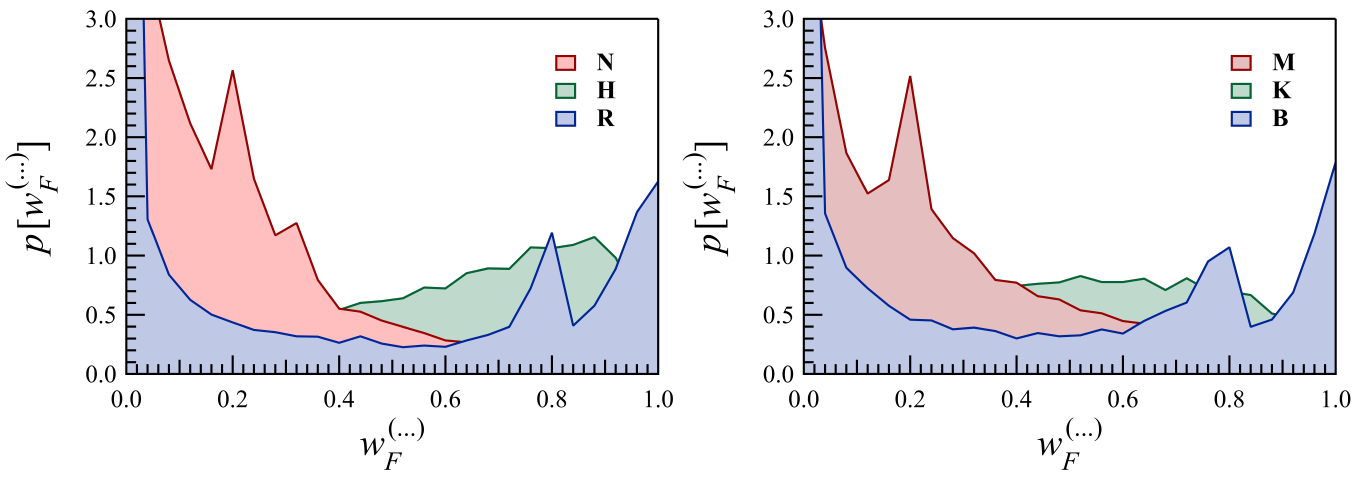}
    \caption{Probability density function (PDFs) of the relative weights $w_F^{(...)}$ for the three components ($\mathbf N$ and $\mathbf M$ elongation/stretching; $\mathbf H$ and $\mathbf K$ shear; $\mathbf R$ and $\mathbf B$ rigid body rotation) of the triple decomposition of the velocity field (left panel) and magnetic field (right panel) gradient tensors for the second magnetosheath period (MSH2).}
    \label{fig:05}
\end{figure}

Figures \ref{fig:04} and \ref{fig:05} show the probability density functions (PDFs) of the relative weights $w_F^{(...)}$ for the three components ($\mathbf N$ and $\mathbf M$ elongation/stretching; $\mathbf H$ and $\mathbf K$ shear; $\mathbf R$ and $\mathbf B$ rigid body rotation) of the triple decomposition of the velocity and magnetic field gradient tensors relative to the first solar wind interval (SW1) and the second magnetosheath interval (MSH2), respectively. Some notable differences can be observed in the two different intervals. For what regards the velocity field in the MSH interval, the shear contribution tends to be dominant with respect to solid body rotation and  elongation/stretching terms. This seems also to be true for the magnetic field. Conversely, in the solar wind interval elongation/stretching and shear terms tend to be the most relevant ones. We remark that in both the intervals the solid body rotation is less relevant, due to the fact that the relevance of this term is associated only to a fraction of the observed structures, i.e., the structures above the discriminant line $\Delta =0$ in the $(R,Q)$ plane of the invariants.

\begin{figure}
    \centering
    \includegraphics[width=0.99\linewidth]{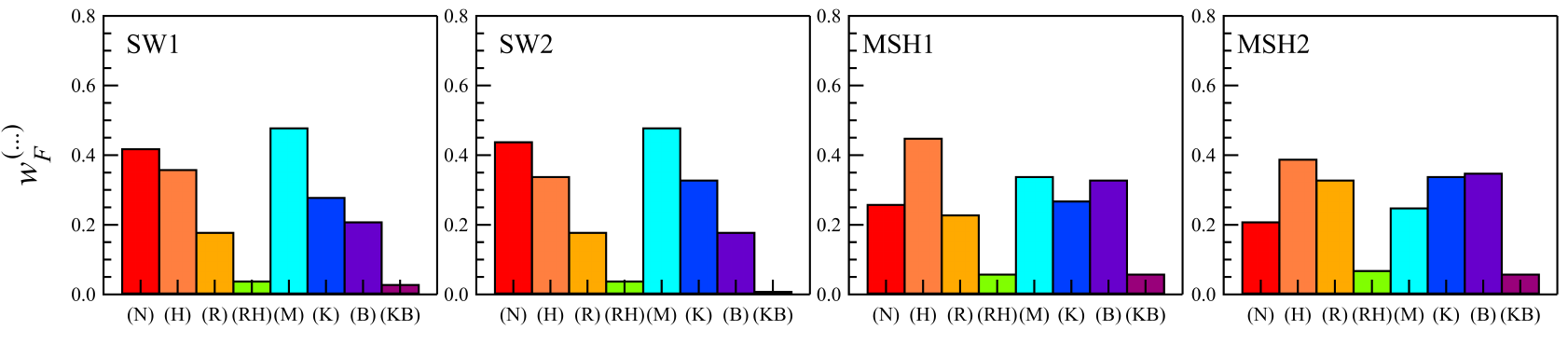}
    \caption{The average values $w_F^{(...)}$ of the relative contribution for the different terms for each of the four studied intervals.}
    \label{fig:05a}
\end{figure}

Table \ref{tab2} and Figure \ref{fig:05a} reports the average values of the relative contribution for the different quantities, including the cross-terms $2RH$ and $2KB$ for each of the four studied intervals. As already noted in the distributions reported in Figures \ref{fig:04} and \ref{fig:05} the elongation/stretching term is essentially dominant in the solar wind while the shear term is really the dominant one in the magnetosheath. This is true for both the velocity and the magnetic field. However, what emerges from the analysis of the average relative contributions is the small relevance of the cross-terms, whose relative weights are $\leq 6-7$\%. This result is completely different from the one by \citet{das2020revisiting} for hydrodynamic turbulence simulations. Indeed, they found $2RH \simeq 13$\%. 

Recalling that $2RH = 2\phi s_3$, the fact that the observed values are really small can be due to two possible reasons:
\begin{itemize}
    \item[i)] rigid body rotation is weak,
    \item[ii)] shear happens not to lie in the rotation plane.
\end{itemize}

The most physically plausible explanation is that, being a non-zero magnetic field in the studied regions, turbulence is anisotropic so to organize its shear structures preferentially away from the local rotation axis. Indeed, if the rotation axes preferentially align along the mean magnetic field, $\mathbf B_0$, and shear is essentially confined to the plane perpendicular to $\mathbf B_0$ (a condition consistent with highly anisotropic and quasi-2D nature of turbulent fluctuations) then geometrically the shear component $s_3$ that ends up in the rotation plane, could be systematically suppressed. This is  not because shear or rotation are individually weak, but because the anisotropy forces them into a geometric configuration where their specific cross-term vanishes, even while shear ($H^2$ and $K^2$) and rotation ($R^2$ and $B^2$) terms individually remain sizable as clearly shown in Table \ref{tab2}.

\begin{table}[]
    \centering
    \begin{tabular}{c|c|c|c|c}
         Term & $w_F^{(SW1)}$ & $w_F^{(SW2)}$ & $w_F^{(MSH1)}$ & $w_F^{(MSH2)}$\\
        \\
        \hline
        $\mathbf N$ & 0.42 & 0.44 & 0.26 & 0.21  \\
        $\mathbf H$ & 0.36 & 0.34 & 0.45 & 0.39  \\
        $\mathbf R$ & 0.18 & 0.18 & 0.23 & 0.33  \\
        $2\mathbf R\mathbf H$ & 0.04 & 0.04 & 0.06 & 0.07  \\
        $\mathbf M$ & 0.48 & 0.48 & 0.34 & 0.25  \\
        $\mathbf K$ & 0.28 & 0.33 & 0.27 & 0.34  \\
        $\mathbf B$ & 0.21 & 0.18 & 0.33 & 0.35  \\
        $2\mathbf K\mathbf B$ & 0.03 & 0.01 & 0.06 & 0.06 
    \end{tabular}
    \caption{The relative weight $w_F^{(...)}$ of the different terms of the triple decomposition of the velocity and magnetic field gradient tensors for the different selected periods.}
    \label{tab2}
\end{table}

In Fig.\ref{fig:05a}, we report a sort of energy balance for different motions in the four different intervals studied. In this figure it is possible to observe that in solar wind there is a predominance of energy distributed in the elongation/stretching motions and this is true for both velocity and magnetic field. Another important point worth emphasizing in the solar wind intervals is that the rigid body rotation motions are those associated with the least amount of energy; once again, this is observed for both the velocity field and the magnetic field. In contrast, if we look at the panels related to the magnetosheath intervals, we immediately note that there is a predominance for shear motions in the energy distribution for the velocity field and an almost uniform distribution of energy in the case of the magnetic field with small differences between an interval and the other one. The key point we wish to emphasise is that, within the magnetosheath, the proportion of energy associated with rigid-body rotational motions has increased, both in the case of the velocity field and in the case of the magnetic field. The varying distribution of energy across different regions of interplanetary space helps us to understand how these regions are governed by different dominant physical mechanisms. It is hardly surprising to recognise that the solar wind consists of a plasma that is freer to flow, whilst the magnetosheath contains a plasma subject to greater rotational motion, given that, amongst other factors, the plasma in the magnetosheath is affected by the physical barrier presented by the Earth.

\begin{figure}
    \centering
    \includegraphics[width=0.9\linewidth]{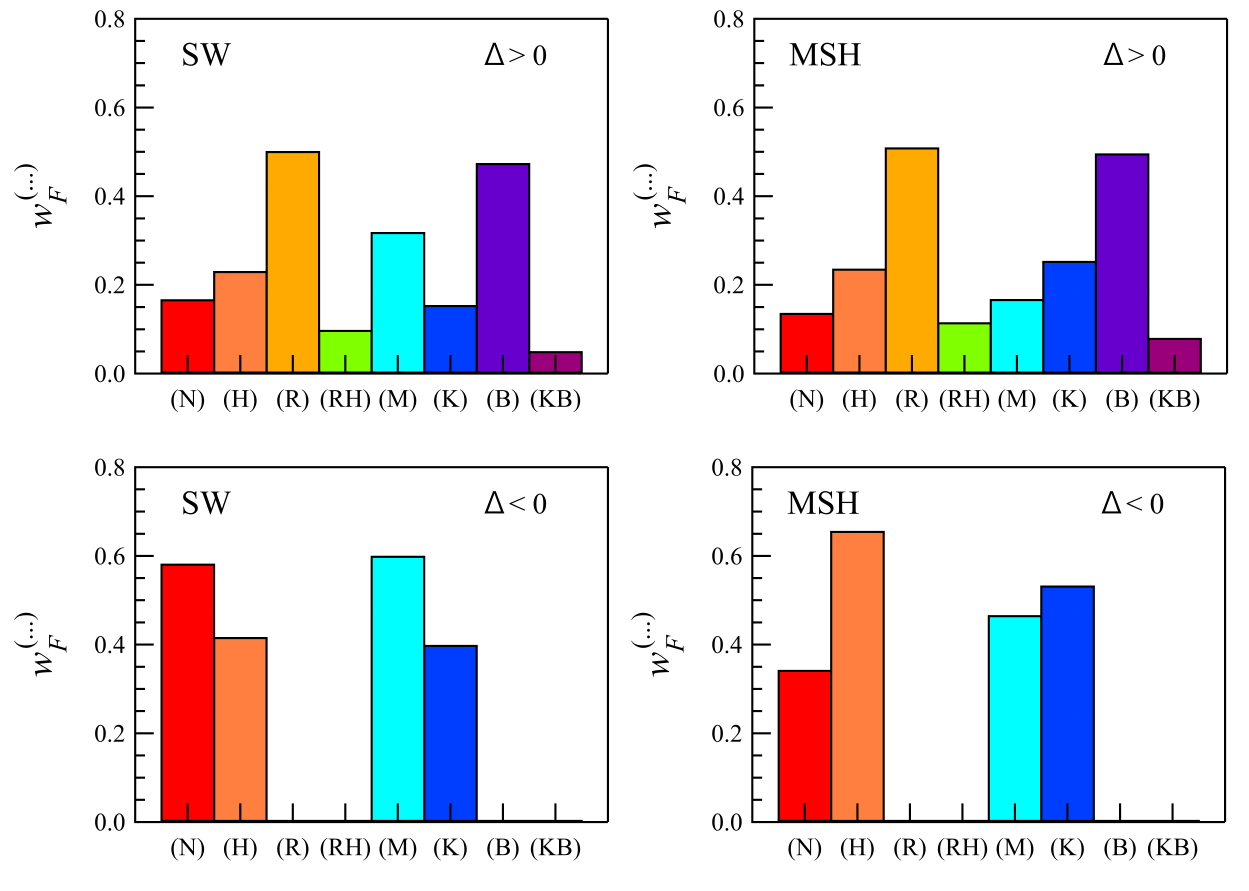}
    \caption{The relative weight $w_F^{(...)}$ of the different terms of the triple decomposition of the velocity and magnetic field gradient tensors for the regions of the $(R,Q)$ plane above ($\Delta >0$) and below ($\Delta <0$) the discriminant line in the case of solar wind intervals (left panels) and magnetosheath intervals (right panels), respectively. Here, $RH$ and $KB$ stand for the cross-terms.}
    \label{fig:06}
\end{figure}

As we noticed in the distributions of the Fr\"obenius norms of the three terms, because the solid body rotation terms is confined above the discriminant line, $\Delta =0$, in the $(R,Q)$-plane, to better quantify the role of this term we compute the relative weights of the terms above and below the discriminant line. The results are shown in Figure \ref{fig:06}, where, for convenience, we have grouped the different regions into two groups, solar wind (SW) and magnetosheath (MSH). Significant differences are observed between the region above the discriminant line associated with ingoing/outgoing spiral saddles and that below the discriminant line, where the structures are mainly tube-like/sheet-like.

In detail, we see how in the case of solar wind the dominant terms for the velocity field are body rotation ($\mathbf{R}$) and, secondarily, shear ($\mathbf{H}$) above the discriminant line and elongation/stretching ($\mathbf{N}$) and shear ($\mathbf{H}$) below the discriminant line where no rigid-body rotation occurs. For what regards the magnetic field the situation is quite similar because the rigid rotation ($\mathbf{B}$) plays is the dominant term, but contrary to the velocity field case we have that the second more important term is the shear ($\mathbf{M}$). Similarly in the magnetosheath the situation above the discriminant line for both the magnetic and velocity field is analogous to the solar wind velocity field but in the second case we can observe that the second more important term is the elongation/stretching ($\mathbf{K}$), making the magnetic case very similar to the velocity one.  An quite different situation is found below the discriminant line, where shear is now dominant for the velocity field while a small prevalence of elongation/stratching can be observed for the magnetic field.

The previous analysis has been performed considering the relevance of the triple decomposition terms based on the squared Fr\"obenius norm (which can be considered a kind of energetic measure). However, we do not know what is the most probable morphological structure dominating the different regions. This can be estimated by computing the persistence of the terms corresponding to the occurrence of maximum local value of the Fr\"obenius norm.

\begin{figure}
    \centering
    \includegraphics[width=0.95\linewidth]{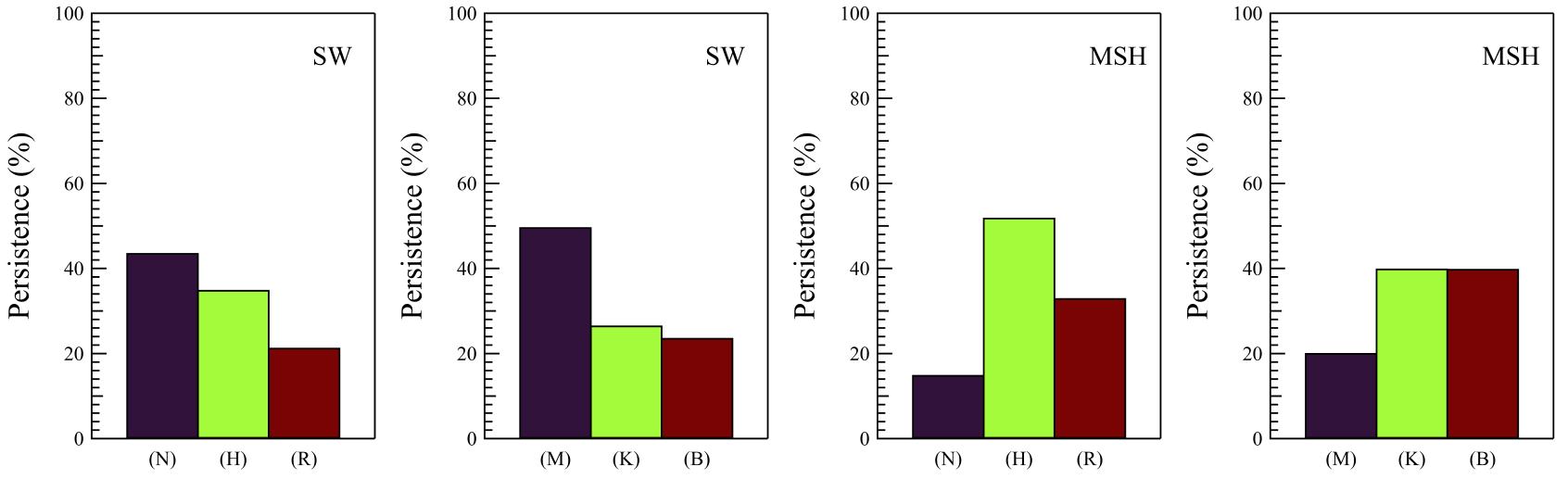}
    \caption{Relative occurrences of time intervals dominated by one of the three motion contributions in the different regions.}
    \label{fig:07}
\end{figure}

Figure \ref{fig:07} reports the relative occurrences of time intervals dominated by one of the three motion terms of the triple decomposition in the different regions. A clear difference is observed between the two regions. This result agrees with what reported in Table \ref{tab2}, again stressing the different morphological character of the turbulence in the two investigated regions. It can be observed that in the magnetosheath there is a significantly higher occurrence of shear-dominated events. This holds true for the velocity, whereas for the magnetic field, we observed a sort of balance between the occurrence of shear motions and rigid-body rotational motions. This discrepancy between the two vector fields could suggest that, while the velocity field is strongly affected by the large-scale flow gradients and boundary interactions, the magnetic field maintains a more complex structure that is potentially influenced by local currents and reconnection processes. In contrast, in the solar wind data a prevalence of strain-dominated events can be observed, although, for the velocity field, there is no a large difference in percentage between the occurrences of strain-dominated and shear-dominated time steps. On the other hand, the prevalence of strain-dominated events is higher for the magnetic field than for the velocity field. This behavior could suggest that the magnetic field is more sensitive to local compressions and expansions within the solar wind, such as tangential discontinuities. These structures enhance the diagonal elements of the gradient tensor, resulting in a prevalence of strain over rotational or shear motions.

\section{Discussions and Conclusions}

The analysis of the results of the triple decomposition of gradient tensors reveals a stark contrast between the two investigated plasma regimes. 

In the magnetosheath region, the dynamics is overwhelmingly dominated by shear, which accounts for a very high percentage of the total occurrences. This massive prevalence of shear, observed more in velocity than in magnetic fields (where it represents, in any case, a dominant behavior), is a clear signature of the intense velocity gradients generated as the plasma is forced to divert around the magnetospheric obstacle. The fact that this contribution is significantly higher in the velocity field further highlights the role of fluid-dynamic interaction with the boundary as the primary driver of this region's turbulence. We have to highlight that for the magnetic field we observe a very important contribution of the rigid-body rotation that is so close to the strain one.

In contrast, the solar wind exhibits a different structural behavior, where strain becomes the predominant contribution. This is consistent with the radial expansion of the solar wind, which induces a continuous stretching of the plasma and its 'frozen-in' magnetic field. However, unlike the magnetosheath, the solar wind shows a much narrower gap between strain and shear occurrences, suggesting a more balanced competition between expansion and local fluctuations. 

Ultimately, these results demonstrate that while the solar wind is shaped by spherical expansion, the magnetosheath is an environment almost entirely defined by boundary-induced shear flow.

Regarding the energy of individual terms of the triple decomposition, it can be observed that magnetosheath data exhibit a very high percentage of energy in rigid-body rotation motions. This is consistent for both the velocity and magnetic fields. In general, magnetosheath data are characterized by a magnetic field in which the rotational energy is very similar to the corresponding energy in the velocity field. Conversely, for solar wind data we observe a much more prominent role for the contribution of strain within the energy distribution. Furthermore, in the solar wind, the magnetic field shows a lower percentage of shear energy and an higher percentage of strain energy compared to the velocity field. It is possible to observe that the magnetic rotational energy is higher than the kinetic rotation one, and both in solar wind data that in magnetosheath one, one can observe similar values for the percentage of rotational energy for velocity and magnetic field.

The observed energy distribution highlights the fundamental differences in the turbulent cascade and driving mechanisms between the magnetosheath and the pristine solar wind. The magnetosheath is a downstream region shaped by the processing of the solar wind plasma through the planetary bow shock. This crossing acts as a major source of vorticity and current sheet formation, effectively converting the bulk kinetic energy of the solar wind into localized fluid shears and intense magnetic fluctuations. Consequently, the high percentage of energy found in shear and rotational motions within the magnetosheath is a direct reflection of a fully developed, strongly driven, and highly intermittent MHD turbulent regime. In this environment, the magnetic field energy displays a stronger rotational component compared to the velocity field, which is consistent with the typical behavior of plasma turbulence at sub-ion scales, where magnetic field line twisting and localized current sheets dominate over fluid velocity vortices.

In contrast, the solar wind data show a more prominent role for strain within the energy distribution. This indicates that the unshocked solar wind plasma is governed by an expansion dominated regime, where global velocity gradients and steady stream-stream interactions continuously stretch and deform the plasma structures lineally, rather than twisting them into compact rotational vortices. Although the magnetic rotational energy remains higher than its kinetic counterpart due to the intrinsic nature of Alfvénic and compressive fluctuations in space plasmas, the overall lower percentage of rotational energy in the solar wind, for both velocity and magnetic field, confirms that its turbulent state is less isotropic and less fragmented than the highly compressed, shock-processed magnetosheath plasma.

Another important point deals with the cross-terms. The evaluated cross-terms ($2RH$ and $2KB$) were found to contribute negligibly ($\le 7\%$) across all studied intervals. This contrasts sharply with hydrodynamic turbulence and suggests that background-magnetic-field-induced anisotropy organizes shear structures perpendicular to the primary rotation axis. Indeed, while hydrodynamic DNS turbulence \citep{das2020revisiting} exhibits a strong coupling ($2RH \approx 13\%$) between shear and rotation planes, space plasma environments demonstrate that $2RH$ and $2KB$ are negligible ($\le 7\%$) across both solar wind and magnetosheath regimes. The vanishingly small value of cross-terms in space plasmas is not caused by weak shear or weak rotation. Instead, it reveals a fundamental geometric ordering imposed by plasma anisotropy. In detail, 
\begin{enumerate}
    \item the presence of a mean background magnetic field $\mathbf{B}_0$ breaks spatial isotropy, so that fluctuations naturally organize into strongly anisotropic structures, where field-aligned gradients differ sharply from perpendicular gradients ($\mathbf{k}_\perp \gg \mathbf{k}_\parallel$);
    \item rotation axes ($\boldsymbol{\phi}$) tend to align preferentially along the local mean magnetic field $\mathbf{B}_0$, and intense shearing motions are confined to the plane perpendicular to $\mathbf{B}_0$;
    \item because the shear dynamics are geometrically constrained to the perpendicular plane, the shear component parallel to/inside the rotation plane ($s_3$) is systematically suppressed ($s_3 \approx 0$).
\end{enumerate}
Thus, in magnetized space plasmas, shear and rotation rarely ``contaminate" one another within the same plane. Turbulence naturally segregates pure rigid rotation (coherent flux ropes/vortices) and pure planar shear (current sheets/shear layers) into geometrically distinct orientations.

In conclusion, we have, here, extended the Triple Decomposition of gradient tensors—originally developed for hydrodynamic velocity fields—to magnetohydrodynamic (MHD) space plasma flows. By applying a real Schur decomposition to high-resolution multi-spacecraft measurements from the NASA-MMS mission, we isolated three kinematically distinct deformation modes for both the velocity ($\mathbf{\nabla v}$) and magnetic ($\mathbf{\nabla B}$) field gradient tensors: irrotational normal strain/stretching ($N$ and $M$), residual pure shear ($H$ and $K$), and true rigid-body rotation ($R$ and $B$).

In summary, the Triple Decomposition framework provides an objective metric for identifying coherent plasma structures and separating pure shearing deformation from genuine rigid-body vortex cores. This methodology offers a powerful diagnostic for future multi-spacecraft missions (e.g., ESA-Plasma Observatory, NASA-HelioSwarm) to investigate cross-scale turbulent energy dissipation, current sheet formation, and magnetic reconnection in space plasmas. The main aim of this work is to highlight how a standard tensor decomposition may not be sufficient to investigate the motions occurring within the space plasma. Indeed, by applying the triple decomposition, it can be seen that regions characterised by $\Delta >0$ are not associated solely with solid-body rotation motions, which means that vorticity differs from this type of motion. Vorticity and current, understood as curls of the velocity field and the magnetic field, consist of a combination of solid-body rotational motions and phenomena of shear and elongation/stretching; only the use of a triple decomposition enables us to separate the contributions to the turbulent motion of space plasma as accurately as possible.

%% Please use the acknowledgment and contribution environments. This will 
%% be anonomyized when the "anonymous" style option is used. 
\begin{acknowledgments}
We acknowledge the P.Is. of FGM and FPI experiments onboard of the NASA-MMS mission and the CDAWeb of the GSFC-SPD (https://cdaweb.gsfc.nasa.gov/) for making freely available data used in this work.
This work is supported by Agenzia Spaziale Italiana (ASI) under the agreement ASI-INAF ”Attività di Fase A per la missione Plasma Observatory" F83C24000690001. 
\end{acknowledgments}

\begin{contribution}
%%This section gives authors the space to recognize author contributions. The text inside this environment is NOT counted towards the total word quanta. At a minimum, manuscripts are expected to include this text:

Conceptualization: VQ; 
Data curation: GC;
Formal analysis and Investigation: VQ \& GC;
Writing – original draft: VQ \& GC;
Writing – review \& editing: All; 
Funding Acquisition: MFM

%% But authors are expected to provide more specific details, e.g. 
%%
%%SC was responsible for writing and submitting the manuscript.
%%WWM came up with the initial research concept and edited the manuscript.
%%OTS obtained the funding and edited the manuscript.
%%EBF provided the formal analysis and validation. He also edited the manuscript.
%%GEH Supervised the undergraduates, wrote the software and administers the project github and Zenodo repositories.
%%
%% Authors can use the Contributor Role Taxonomy (CRediT) at
%% for ideas on how write a good statement tailored to their needs.

\end{contribution}

\bibliography{Quattrociocchi.bib}{}
\bibliographystyle{aasjournalv7}



\end{document}